\documentclass{JHEP}
\usepackage{amsmath,amssymb}
\usepackage{epsfig}

\newcommand{\eq}{\begin{equation}}
\newcommand{\feq}{\end{equation}}
\newcommand{\eqn}{\begin{eqnarray}}
\newcommand{\feqn}{\end{eqnarray}}
\newcommand{\arr}{\begin{eqnarray*}}
\newcommand{\farr}{\end{eqnarray*}}

\title{Causality Violation and Naked Time Machines in AdS$_5$}
\author{Marco M.~Caldarelli \\
Dipartimento di Fisica dell'Universit\`a di Trento and \\
INFN, Gruppo Collegato di Trento, Via Sommarive 14,
38050 Povo (TN), Italy.\\
E-mail: \email{caldarel@science.unitn.it}}
\author{Dietmar Klemm \\
Dipartimento di Fisica dell'Universit\`a di Milano and \\
INFN, Sezione di Milano, Via Celoria 16, 20133 Milano, Italy.\\
E-mail: \email{dietmar.klemm@mi.infn.it}}
\author{Wafic A.~Sabra \\
Center for Advanced Mathematical Sciences (CAMS) and\\
Physics Department, American University of Beirut, Lebanon.\\
E-mail: \email{ws00@aub.edu.lb}}
\preprint{UTF-443\\
IFUM-682-FT\\
CAMS/01-02\\
hep-th/0103133}

\abstract{We study supersymmetric charged rotating black holes in AdS$_5$,
and show that closed timelike curves occur outside the event horizon.
Also upon lifting to rotating D3 brane solutions of type IIB
supergravity in ten dimensions, closed timelike curves are still present.
We believe that these causal anomalies correspond to loss of
unitarity in the dual ${\cal N}=4$, $D=4$ super Yang-Mills theory,
i.~e.~the chronology protection conjecture in the AdS bulk is related
to unitarity bounds in the boundary CFT.
We show that no charged or uncharged geodesic can penetrate the horizon, so
that the exterior region is geodesically complete. These results still hold
true in the quantum case, i.~e.~the total absorption cross section for
Klein-Gordon scalars propagating in the black hole background is zero. This
suggests that the effective temperature is zero instead of assuming the
naively found imaginary value.}


\begin{document}

\section{Introduction}

It is well-known that supersymmetry does not exclude the existence of closed
timelike curves. Simple examples are flat space with periodically
identified time and anti-de~Sitter space. However, these spacetimes are not
simply connected, and one can avoid CTCs by passing to the universal covering.
One might therefore assume that in simply connected supersymmetric spaces
closed timelike curves do not exist. This however turned out not to be the
case, for example the BMPV black hole \cite{bmpv,kallosh}, which is a BPS
solution of $D=5$, $\mathcal{N}=2$ supergravity\footnote{Cf.~also
\cite{chamsabra} for a generalization of the BMPV solution to the case of
$D=5$, $\mathcal{N}=2$ supergravity coupled to vector multiplets.} with
trivial fundamental group, admits CTCs. These black holes are characterized by
a charge parameter $q$ and a rotation parameter $a$. For $a^{2}<q^{3}$
(hereafter referred to as the ``under-rotating'' case), CTCs occur only in the
region beyond the horizon, and thus an external observer cannot use them to
construct a time machine. This is quite similar to the case of the Kerr black
hole. However, in the ``over-rotating'' case $a^{2}>q^{3}$, CTCs are present
in the exterior region outside the event horizon \cite{gauntlett}. This
happens in spite of supersymmetry and the presence of a matter stress tensor
satisfying the dominant energy condition. Various aspects of these naked time
machines have been extensively analyzed in \cite{gibbherd,herd}. Among other
things, the authors of \cite{gibbherd} showed that causal geodesics cannot
penetrate the horizon for the over-rotating solution (repulson-like
behaviour), and so the exterior region is geodesically complete (with respect
to causal geodesics). Furthermore, in \cite{herd} it was shown that upon
lifting of the over-rotating BMPV black hole to a solution of type IIB string
theory and passing to the universal covering space, the causal anomalies
disappear. The rotating BMPV black hole, which corresponds to a
D1-D5-Brinkmann wave system in type IIB string theory \cite{herd}, admits a
dual description in terms of an $\mathcal{N}=4$ two-dimensional superconformal
field theory. The causality bound $a^{2}=q^{3} $ is then equivalent to the
unitarity bound in this CFT \cite{herd}, obtained by requiring unitary
representations of the superconformal algebra, which implies an inequality
involving the central charge, conformal weight and R-charge.\newline Recently,
the generalization of the BMPV solution to the case of $D=5$, $\mathcal{N}=2$
gauged supergravity has been found in \cite{ks1,ks2}. These black holes, which
also preserve four supercharges, were shown to suffer from CTCs outside the
horizon for all parameter values, as soon as rotation is turned on. Due to
these causal anomalies, the naively computed Hawking temperature and the
Bekenstein-Hawking entropy are imaginary. According to the AdS/CFT
correspondence \cite{adscft}, these black hole solutions, which asymptotically
approach anti-de~Sitter space, should be dual to $\mathcal{N}=4$, $D=4$ super
Yang-Mills theory in the presence of $R$-charges. Similar to the ungauged
case, where the appearance of CTCs was shown to be related to unitarity
violation in a two-dimensional CFT, we thus expect that the causal anomalies
occurring in the charged rotating AdS black holes correspond to loss of
unitarity in $\mathcal{N}=4$, $D=4$ super Yang-Mills theory. One has thus a
relation between macroscopic causality in the AdS bulk and microscopic
unitarity in the boundary CFT.\newline In this paper, we will be concerned
with a detailed study of the rotating AdS black holes found in \cite{ks1,ks2}
and their causal anomalies. Our work is organized as follows. In sections
\ref{sect:bhsol} and \ref{sect:geomprop}, we review the solution and analyze
its geometric properties. In section \ref{sect:globstr} the throat geometry
describing the near-horizon limit of the extremal solutions is determined. In
\ref{sect:geodesics} it is shown that no charged or uncharged geodesic can
penetrate the horizon, implying that the exterior region is geodesically
complete. This behaviour still holds true in the quantum case, that is the
absorption cross section of the hole for Klein-Gordon scalars is zero. This
suggests that the effective Hawking temperature is zero instead of assuming
the naively found imaginary value. The results obtained in section
\ref{sect:geodesics} are thus very similar to those of the ungauged case that
was studied in \cite{gibbherd,herd}. In \ref{sect:vectmult} our results are
generalized to the case when the gauged supergravity theory is coupled to
vector multiplets. The solution of the $STU=1$ model is lifted to a solution
of type IIB supergravity in ten dimensions in section \ref{sect:lift},
yielding a rotating D3 brane wrapping $\mathcal{S}^{3}$. It is then shown that
the original CTCs present in five dimensions disappear upon lifting, but new
CTCs show up. Therefore, in contrast to the BMPV black holes, we have no
resolution of causal anomalies in higher dimensions. We conclude in
\ref{sect:finrem} with some final remarks.

\section{The black hole solution}

\label{sect:bhsol}

We consider Einstein-Maxwell theory in five dimensions with a negative
cosmological constant and a Chern-Simons term for the abelian gauge field. The
action is given by
\begin{equation}
S=\frac{1}{16\pi G_{5}}\int d^{5}x\sqrt{-g}\left(  R-2\Lambda-\frac{1}%
{12}F_{\mu\nu}F^{\mu\nu}\right)  +\frac{1}{16\pi G_{5}}\int d^{5}x\frac
{1}{108}\epsilon^{\mu\nu\rho\sigma\lambda}F_{\mu\nu}F_{\rho\sigma}A_{\lambda},
\label{EMaction}%
\end{equation}
where $R$ is the scalar curvature, $F_{\mu\nu}=\partial_{\mu}A_{\nu}%
-\partial_{\nu}A_{\mu}$ the abelian field-strength tensor, $G_{5}$ denotes the
five-dimensional Newton constant and $\Lambda=-6g^{2}$ the cosmological
constant. (\ref{EMaction}) is the bosonic truncation of the gauged
five-dimensional $\mathcal{N}=2$ pure supergravity theory.\newline In
\cite{ks1}, a charged rotating solution of this theory was found. Its metric
reads
\begin{equation}
ds^{2}=-g^{2}r^{2}\ dt^{2}-\frac{1}{r^{4}}\left[  \left(  r^{2}-q\right)
dt-a\sin^{2}\theta\ d\phi+a\cos^{2}\theta\ d\psi\right]  ^{2}+\frac{dr^{2}%
}{V(r)}+r^{2}\ d\Omega_{3}^{2}, \label{KSmetric}%
\end{equation}
where
\begin{equation}
V(r)=\left(  1-\frac{q}{r^{2}}\right)  ^{2}+g^{2}r^{2}-\frac{g^{2}a^{2}}%
{r^{4}},
\end{equation}
and the gauge fields are given by
\begin{equation}
A_{\phi}=-\frac{3a}{r^{2}}\sin^{2}\theta,\qquad A_{\psi}=\frac{3a}{r^{2}}%
\cos^{2}\theta,\qquad A_{t}=3\left(  1-\frac{q}{r^{2}}\right)  .
\end{equation}
Here, $d\Omega_{3}^{2}$ denotes the standard metric on the unit three-sphere,
\begin{equation}
d\Omega_{3}^{2}=d\theta^{2}+\sin^{2}\theta\ d\phi^{2}+\cos^{2}\theta
\ d\psi^{2},
\end{equation}
with the angles $\theta$, $\phi$ and $\psi$ parametrizing $\mathcal{S}^{3} $
ranging in $\theta\in\lbrack0,\pi/2]$, $\phi\in\lbrack0,2\pi\lbrack$, $\psi
\in\lbrack0,2\pi\lbrack$.\newline For convenience, we define two other
functions, which will be of use in the following, namely
\begin{equation}
\Delta(r)\equiv1-\frac{q}{r^{2}}\,,\qquad\Delta_{L}(r)\equiv1-\frac{a^{2}%
}{r^{6}}\,.
\end{equation}
Using the diffeomorphism $\mathcal{S}^{3}\cong SU(2)$, one can parametrize
the three-sphere with the Euler parameters $(\alpha,\beta,\gamma)$ of $SU(2)$,
related to the angular variables $\theta$, $\phi$ and $\psi$ through the
transformation
\begin{equation}
\alpha=\psi+\phi\,,\quad\beta=2\theta\,,\quad\gamma=\psi-\phi.
\end{equation}
In these coordinates, the metric of the three-sphere takes the form
\begin{equation}
d\Omega_{3}^{2}=\frac{1}{4}\left(  d\alpha^{2}+d\beta^{2}+d\gamma^{2}%
+2\cos\beta\ d\alpha\ d\gamma\right)  \,,
\end{equation}
with $\alpha\in\lbrack0,2\pi\lbrack$, $\beta\in\lbrack0,\pi]$ and $\gamma
\in\lbrack0,4\pi\lbrack$. The isometry group of the three-sphere $SO(4)\cong
SU(2)_{L}\times SU(2)_{R}$ consists of two copies of the $SU(2)$ group. The
left-invariant vector fields $\xi^{R}$ generating the right translations are
\begin{align}
&  \xi_{1}^{R}=\cos\gamma\,\mathrm{cosec}\,\beta\ \partial_{\alpha}-\sin
\gamma\ \partial_{\beta}-\cos\gamma\cot\beta\ \partial_{\gamma}\,,\nonumber\\
&  \xi_{2}^{R}=\sin\gamma\,\mathrm{cosec}\,\beta\ \partial_{\alpha}+\cos
\gamma\ \partial_{\beta}-\sin\gamma\cot\beta\ \partial_{\gamma}\,,
\label{xir}\\
&  \xi_{3}^{R}=\partial_{\gamma}\,.\nonumber
\end{align}
whereas the right-invariant vector fields generating the left translations
read
\begin{align}
&  \xi_{1}^{L}=-\sin\alpha\cot\beta\ \partial_{\alpha}+\cos\alpha
\ \partial_{\beta}+\sin\alpha\,\mathrm{cosec}\,\beta\ \partial_{\gamma
}\,,\nonumber\\
&  \xi_{2}^{L}=-\cos\alpha\cot\beta\ \partial_{\alpha}-\sin\alpha
\ \partial_{\beta}+\cos\alpha\,\mathrm{cosec}\,\beta\ \partial_{\gamma
}\,,\label{xil}\\
&  \xi_{3}^{L}=\partial_{\alpha}\,.\nonumber
\end{align}
They satisfy the commutation relations
\begin{equation}
\left[  \xi_{a}^{R},\xi_{b}^{R}\right]  =-\epsilon_{abc}\xi_{c}^{R}%
\ ,\qquad\left[  \xi_{a}^{L},\xi_{b}^{L}\right]  =\epsilon_{abc}\xi_{c}%
^{L}\ ,\qquad\left[  \xi_{a}^{R},\xi_{b}^{L}\right]  =0\ .
\end{equation}
Introducing the left-invariant one-forms $\sigma_{i}$ ($i=1,2,3$), dual to the
left-invariant vector fields $\xi_{a}^{R}$ in the sense that $(\xi_{a}%
^{R},\sigma_{b})=\delta_{ab}$,
\begin{eqnarray}
&&\sigma_{1}=-\sin\gamma\ d\beta+\cos\gamma\sin\beta\ d\alpha\,,\nonumber\\
&&\sigma_{2}=\cos\gamma\ d\beta+\sin\gamma\sin\beta\ d\alpha\,,\\
&&\sigma_{3}=d\gamma+\cos\beta\ d\alpha\,,\nonumber
\label{sigma}\end{eqnarray}
the metric of the three sphere can be written as
\begin{equation}
d\Omega_{3}^{2}=\frac{1}{4}\left(  \sigma_{1}^{2}+\sigma_{2}^{2}+\sigma
_{3}^{2}\right)  .
\end{equation}
The left-invariant one-forms satisfy $d\sigma_{1}=\sigma_{2}\wedge\sigma_{3}$
(together with its cyclic permutations). Rewriting the black hole metric
(\ref{KSmetric}) in terms of the one-forms $\sigma^{i}$, we obtain
\begin{equation}
ds^{2}=-g^{2}r^{2}\ dt^{2}-\Delta^{2}(r)\left(  dt+\frac{a}{2r^{2}\Delta
(r)}\ \sigma_{3}\right)  ^{2}+\frac{dr^{2}}{V(r)}+\frac{r^{2}}{4}\left(
\sigma_{1}^{2}+\sigma_{2}^{2}+\sigma_{3}^{2}\right)  , \label{metric:f}%
\end{equation}
and the gauge-field 1-form is given by
\begin{equation}
A=3\Delta(r)\ dt+\frac{3a}{2r^{2}}\ \sigma_{3}.
\end{equation}

\section{Geometric properties of the black hole spacetime}

\label{sect:geomprop}

\subsection{``Horizons''}

The horizons of the black hole are located at the zeroes of the function
$V(r)$. Using the dimensionless parameters $\alpha=g^{3}a$, $\rho=g^{2}q$ and
$\zeta=gr$, the problem reduces to finding the zeroes of the polynomial
$f(\zeta)$ defined by
\begin{equation}
f(\zeta)=\zeta^{6}+\zeta^{4}-2\rho\zeta^{2}+\rho^{2}-\alpha^{2}.
\end{equation}
If $\rho\leq0$, we have $f^{\prime}(\zeta)=0$ only for $\zeta=0$, and hence
$f(\zeta)$ is an increasing function of its variable. As $f(0)=\rho^{2}%
-\alpha^{2} $, for $|\rho|>|\alpha|$ we have a naked singularity, for
$|\rho|=|\alpha|$ a single horizon at $\zeta_{+}=0$, and for $|\rho|<|\alpha|$
a single horizon located at some point $\zeta_{+}>0$, with a spacelike singularity.

The situation is more interesting for $\rho>0$. In this case, the derivative
$f^{\prime}$ has an additional positive root for
\begin{equation}
\zeta^{2} = \bar\zeta^{2} \equiv\frac13\left(  \sqrt{1+6\rho}-1\right)  .
\label{der0}%
\end{equation}
Hence, if $\bar\zeta$ is also a root of $f$, it is a double root and thus the
black hole is extremal. This occurs for $\alpha^{2}=\alpha_{extr}^{2}%
\equiv\bar\zeta^{6}+\bar\zeta^{4}-2\rho\bar\zeta^{2}+\rho^{2}$. The critical
rotation parameter is given by
\begin{equation}
\alpha_{extr}^{2}(\rho)=\rho^{2}+\frac23\rho+\frac2{27}-\frac2{27}\left(  1
+6\rho\right)  ^{3/2}.
\end{equation}
It is easy to verify that $\rho^{2}>\alpha_{extr}^{2}$ holds for $\rho>0$.
Hence, for positive $\rho$, we have the following cases:

\begin{itemize}
\item $\rho^{2}>\alpha_{extr}^{2}(\rho)>\alpha^{2}$ : There is no positive
root, and hence no horizon and therefore we are left with a naked singularity.

\item $\alpha^{2}=\alpha_{extr}^{2}(\rho)$ : There is a double root $\zeta
_{+}>0$ of $f$, and thus the black hole is extremal. The singularity is
timelike, and the horizon is exactly at $\zeta_{+}=\bar{\zeta}$ given by
equation~(\ref{der0}).

\item $\rho^{2}>\alpha^{2}>\alpha_{extr}^{2}(\rho)$ : There are two roots
$\zeta_{+}$ and $\zeta_{-}$, corresponding to an outer event horizon at
$\zeta_{+}$, and an inner Cauchy horizon at $\zeta_{-}$. In this case, the
singularity is spacelike.

\item $\alpha^{2} = \rho^{2}$ : The Cauchy horizon collapses in the
singularity, $\zeta_{-}=0$, leaving a simple root $\zeta_{+}>0$ corresponding
to the bifurcated horizon of the black hole.

\item $\alpha^{2}>\rho^{2}$ : There is a single positive root $\zeta_{+}$, and
the metric describes a black hole with a spacelike singularity hidden by a
bifurcated horizon located at $\zeta_{+}$.
\end{itemize}
The results are summarized in figure~\ref{fig:rhoalpha}, where the
properties of our metric are shown in the $(\alpha,\rho)$-plane.
\FIGURE{
\epsfig{file=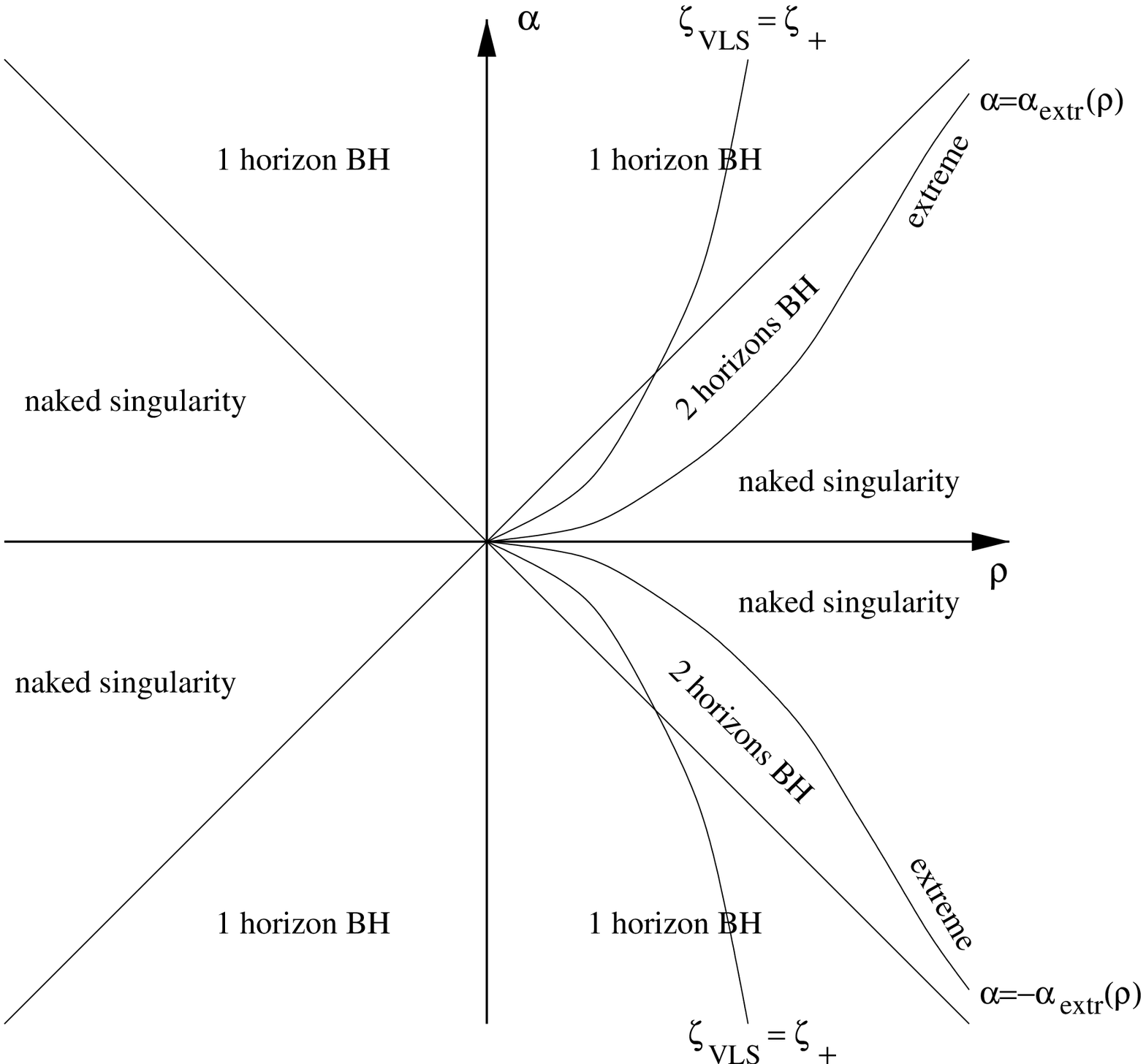,angle=0,width=\linewidth}%
\caption{\textsf{Properties of the metric on the $(\alpha,\rho)$-plane.}}%
\label{fig:rhoalpha}}

The one-parameter subfamily of extreme black holes can be parametrized by the
location $\zeta_{+}$ of the double root of $f$. In this case, $f$ takes the
form
\begin{equation}
f(\zeta)=\left(  \zeta^{2}-\zeta_{+}^{2}\right)  ^{2}\left(  \zeta^{2}%
+2\zeta_{+}^{2}+1\right)  ,
\end{equation}
and the rotation and charge parameters are given by
\begin{equation}
\rho=\frac{1}{2}\left(  3\zeta_{+}^{2}+2\right)  \zeta_{+}^{2}\,,\qquad
\alpha^{2}=\left(  \frac{9}{4}\zeta_{+}^{2}+1\right)  \zeta_{+}^{6}\,.
\end{equation}
With these relations, it is possible to explicitly write the metric components
of the extreme metrics as a function of $\zeta_{+}$.

We finally look for ergospheres. Ergoregions occur when the norm of the
asymptotically timelike Killing vector $\partial_{t}$ becomes positive. But
$(\partial_{t},\partial_{t})=-g^{2}r^{2}-\Delta^{2}(r)<0$ for every $r$, hence
$\partial_{t}$ is always timelike and there are no ergospheres in the manifold
under consideration.

\subsection{Velocity of light surfaces and time machines}

It can be easily seen that our metric allows for closed timelike curves.
For instance, the norm of $\partial_{\gamma}$ is $r^{2}/4-a^{2}%
/(4r^{4})$ and becomes timelike for $r<a^{1/3}$. The integral curves of this
Killing vector being closed circles, we see that for $r$ sufficiently small
there are CTCs.
It is then essential to see whether these curves arise behind
the horizon - as for the Kerr black hole - or if they are outside, thus
yielding a naked time machine undermining causality.

On the boundary between the time machine region and the causal region, the
Killing vector $\partial_{\gamma}$ becomes lightlike. As in \cite{gibbherd} we
shall call this boundary the \emph{velocity of light surface} (VLS); it is a
timelike surface located in $\zeta_{VLS}=\alpha^{1/3}$. If this surface is not
hidden by a horizon, our spacetime is a naked time machine.

Rearranging terms in the polynomial $f$, we obtain $f(\zeta)=\zeta^{6}%
-\alpha^{2}+(\zeta^{2}-\rho)^{2}$, which implies
\begin{equation}
\zeta_{VLS}=\left|  \alpha\right|  ^{1/3}\geq\zeta_{+} \label{VLS}%
\end{equation}
for every choice of the parameters $(\alpha,\rho)$ giving rise to a horizon.
Hence, the velocity of light surface is always \emph{outside} of the horizon,
and the black hole (\ref{KSmetric}) has naked CTCs for every choice of
parameters\footnote{The appearance of naked CTCs is a general feature of
spinning black holes that are solutions of the vacuum Einstein equations
in any odd number of dimensions, when all the rotation parameters are
nonvanishing, as shown in \cite{Myers:1986un}.}
(this contrasts with its asymptotically flat ($g=0$)
counterpart, where the VLS is hidden by an event horizon for $a^{2} < q^{3}$
\cite{gibbherd}). When equality holds in equation (\ref{VLS}), the VLS
coincides with the horizon, $\zeta_{VLS}=\zeta_{+}$. This occurs for
$f(\zeta_{VLS})=(\alpha^{2/3}-\rho)^{2}=0 $, that is for the parameters of the
black hole satisfying $\alpha^{2}_{VLS}=\rho^{3}$. This curve is shown in
figure~\ref{fig:rhoalpha}.

Analyzing the sign of $(\partial_{t},\partial_{\gamma})$, we see that
$\partial_{\gamma}$ is future pointing for $\alpha>0$, and past pointing for
$\alpha<0$.

\subsection{Symmetries and Killing tensors}

The spacetime (\ref{metric:f}) is stationary, hence $\partial_{t}$ is a
Killing vector. Turning to the isometries of the three-sphere (\ref{xir}),
(\ref{xil}), we see that the rotation in the right $SU(2)$ sector of $SO(4)$
breaks the $\xi_{1}^{R}$ and $\xi_{2}^{R}$ right translations, and hence
breaks the $SU(2)_{R}$ isometry group to $U(1)_{R}$ generated by
$\xi_{3}^{R}$.
On the other side, it can be verified that the left rotation isometry subgroup
$SU(2)_{L}$ remains unbroken, and the vectors $\xi_{i}^{L}$ given in
(\ref{xil}) are Killing vectors of the metric. Thus, the
isometry group of our spacetime is $\mathbb{{}R}\times SU(2)_{L}\times
U(1)_{R}$.\newline In the following sections, we shall study the geodesics of
this metric. The Killing vectors give rise to three conserved charges, which
will show useful in the separation of the angular part of the equations.
Furthermore, a fourth conserved quantity is provided by the
St\"{a}ckel-Killing tensor
\begin{equation}
K^{\mu\nu}=\sum_{i=1}^{3}\xi_{i}^{L}{}^{\mu}\xi_{i}^{L}{}^{\nu}=\sum_{i=1}%
^{3}\xi_{i}^{R}{}^{\mu}\xi_{i}^{R}{}^{\nu}\,.
\end{equation}
This is the Casimir invariant of any of the $SU(2)$ subgroups of the $SO(4)$
rotation group, which coincide in the scalar representation. As a consequence,
breaking $SU(2)_{R}\rightarrow U(1)_{R}$ does not decrease the actual number
of constant of motions, and even in the rotating case the geodesic and wave
equations remain completely separable. We refer the reader to \cite{gibbherd}
for a complete discussion.

\section{Throat geometry of the near-extreme solution}

\label{sect:globstr} To perform the near-extremal, near-horizon limit, it is
useful to start from the canonical form of the metric
\begin{equation}
ds^{2}=-\frac{V(r)}{\Delta_{L}}dt^{2}+\frac{dr^{2}}{V(r)}+\frac{r^{2}}%
{4}\Delta_{L}\left(  \sigma_{3}-\omega\ dt\right)  ^{2}+\frac{r^{2}}{4}\left(
\sigma_{1}^{2}+\sigma_{2}^{2}\right)\,,
\end{equation}
where we have defined
\begin{equation}
\omega=\frac{2a\Delta}{r^{4}\Delta_{L}}\ .
\end{equation}
To obtain a non-singular near-horizon limit, we have to approach the extremal
limit while moving towards the horizon. Following \cite{Caldarelli:2000wc}, we
keep the charge $q$ fixed and parametrize the rotation parameter as
$a=a_{extr}(q)(1+k\epsilon^{2})$, where $k$ is an arbitrary constant, and
$\epsilon$ is the extremality parameter. Let us call $r_{e}$ the location of
the double root of $V(r)$ which develops for $\epsilon=0$. For simplicity, we
define also the rotation parameter at extremality $a_{e}=a_{extr}(q)$ and the
angular velocity at extremality
\begin{equation}
\omega_{e}=\left.  \frac{2a_{e}\Delta}{r^{4}\Delta_{L}}\right|  _{extr}%
=\frac{4a_{e}}{3r_{e}^{4}}.
\end{equation}
In order to take the near-horizon limit as we approach the extremal solution,
we define the new coordinates $(\psi,R,\gamma_{c})$
\begin{equation}
r=r_{e}+\epsilon R\ ,\qquad\psi=\epsilon t\ ,\qquad\gamma_{c}=\gamma
-\omega_{e}t\ ;
\end{equation}
hence $\sigma_{3}$ becomes
\begin{equation}
\sigma_{c}=\sigma_{3}-\omega_{e}\ dt\ .
\end{equation}
One then finally performs the $\epsilon\rightarrow0$ limit. The function
$V(r)$ reads
\begin{equation}
V(r)=\left[  4\left(  1+3g^{2}r_{e}^{2}\right)  \frac{R^{2}}{r_{e}^{2}}%
-\frac{2g^{2}a_{extr}^{2}(r_{e})}{r_{e}^{4}}k\right]  \epsilon^{2}%
+\mathcal{O}\left(  \epsilon^{3}\right)
\end{equation}
and the near horizon metric is
\begin{equation}
ds^{2}=\frac{4f(R)}{9g^{2}r_{e}^{2}}\ d\psi^{2}+\frac{dR^{2}}{f(R)}%
-\frac{9g^{2}r_{e}^{4}}{16}\left(  \sigma_{c}-\frac{16a_{e}}{9g^{2}r_{e}^{7}%
}\ Rd\psi\right)  ^{2}+\frac{r_{e}^{2}}{4}\left(  \sigma_{1}^{2}+\sigma
_{2}^{2}\right)  \ , \label{throatmetr}%
\end{equation}
with
\begin{equation}
f(R)=4\left(  1+3g^{2}r_{e}^{2}\right)  \frac{R^{2}}{r_{e}^{2}}-2k\frac
{g^{2}a_{e}^{2}}{r_{e}^{4}}\ .
\end{equation}
Similar throat solutions (with $k=0$) have been found in
\cite{Zaslavsky:1998uu,Bardeen:1999px} for Kerr black holes and in
\cite{Clement:2001ms,Clement:2001gi} for black holes in five-dimensional
dilaton-axion gravity. To obtain a finite limit for the vector potential, we
have to perform a gauge transformation $A\mapsto A-2dt$. The limit can then be
performed safely, obtaining
\begin{equation}
A=\frac{2R}{r_{e}}\ d\psi+\frac{3a_{e}}{2r_{e}^{2}}\ \sigma_{c}\ .
\label{Athroat}%
\end{equation}
For $k=0$, the throat solution has an enhanced isometry group; the original
$SU(2)_{L}\times U(1)_{R}$ symmetry generated by $\xi_{i}^{L}$ and $\xi
_{3}^{R}$ is still present, but we have now three additional Killing vectors
\begin{align*}
\chi_{1}  &  =\frac{\partial}{\partial\psi}\,,\qquad\qquad\chi_{2}=\psi
\frac{\partial}{\partial\psi}-R\frac{\partial}{\partial R}\,,\\
\chi_{3}  &  =\left[  \frac{\psi^{2}}{2}-\frac{9g^{2}r_{e}^{6}}{128\left(
1+3g^{2}r_{e}^{2}\right)  ^{2}}\frac{1}{R^{2}}\right]  \frac{\partial
}{\partial\psi}-\psi R\frac{\partial}{\partial R}-\frac{a_{e}}{4r_{e}\left(
1+3g^{2}r_{e}^{2}\right)  ^{2}}\frac{1}{R}\frac{\partial}{\partial\gamma_{c}%
}\,,
\end{align*}
which obey the commutation relations of the $SL(2,{}\mathbb{R})$ algebra. As a
consequence, the full isometry group of the throat solution is $SL(2,{}%
{}\mathbb{R})\times SU(2)_{L}\times U(1)_{R}$. A non-vanishing $k$, in
contrast, breaks the $SL(2,{}{}\mathbb{R})$ symmetry group and only $\chi_{1}$
remains a Killing vector. In this case the isometry group is ${}{}%
\mathbb{R}\times SU(2)_{L}\times U(1)_{R}$, as in the full metric.\newline The
near-horizon limit (\ref{throatmetr}), (\ref{Athroat}) with $k=0$ is a special
case of a general class of solutions to (\ref{EMaction}), given by
\begin{align}
ds^{2}  &  =c_{1}^{2}(d\chi^{2}+\sinh^{2}\chi\ d\zeta^{2})+c_{2}^{2}(d\beta
^{2}+\sin^{2}\beta\ d\alpha^{2})-(d\gamma+b_{1}A_{1}+b_{2}A_{2})^{2}%
\,,\nonumber\\
A  &  =a_{1}A_{1}+a_{2}A_{2}\,,\qquad
A_{1}=\cosh\chi\ d\zeta\,,\qquad
A_{2}=\cos\beta\ d\alpha\,. \label{genthroat}
\end{align}
(Note that $dA_{1}$ and $dA_{2}$ are essentially the K\"{a}hler forms on
$H^{2}$ and $S^{2}$ respectively). The constants $a_{i}$, $b_{i}$ and $c_{i}$
are subject to the constraints
\begin{align}
&a_{1}^{2}=6b_{1}^{2}-6c_{1}^{2}+3b_{2}^{2}(c_{1}/c_{2})^{4}\,,\qquad
a_{2}^{2}=6b_{2}^{2}+6c_{2}^{2}+3b_{1}^{2}(c_{2}/c_{1})^{4}\,,\nonumber\\
&12g^{2}=c_{1}^{-2}-c_{2}^{-2}\,,\quad2c_{1}^{2}c_{2}^{2}a_{1}a_{2}+\sqrt
{3}(c_{2}^{4}a_{1}b_{1}+c_{1}^{4}a_{2}b_{2})=0\,,\nonumber
\end{align}
following from the equations of motion of the action (\ref{EMaction}). This
system leaves two parameters undetermined, e.~g.~we can choose $c_{1}^{2}$ and
the ratio $b_{1}/a_{2}$ freely\footnote{In particular, for the choice
$a_{1}=b_{2}=0$, $a_{2}=1/g$, $b_{1}^{2}=c_{1}^{2}=1/(9g^{2})$, $c_{2}%
^{2}=-1/(3g^{2})$, one obtains the solutions considered in \cite{klemmsabra},
of the form $AdS_{3}\times H^{2}$, where the $AdS_{3}$ part is written as an
$S^{1}$ bundle over $H^{2}$. These solutions preserve half of the
supersymmetries.}. It is straightforward to show that the general solution
(\ref{genthroat}) is a homogeneous manifold
$[SO(2,1)\times SU(2)\times U(1)]/[U(1)\times U(1)]$.
It should therefore be interesting to study this
spacetime in the context of holography and coset spaces \cite{strom}.

\section{Maximal extension of the black hole solution}
\label{sect:geodesics}

\subsection{Geodesics of the black hole solution}

The Hamilton-Jacobi equation for the action function $S(x^{\mu})$, describing
the geodesics of the metric (\ref{KSmetric}), reads
\begin{align}
&  \displaystyle-\frac{\Delta_{L}}{V}\left(  \frac{\partial S}{\partial
t}\right)  ^{2}+V\left(  \frac{\partial S}{\partial r}\right)  ^{2}%
-\frac{4a\Delta}{r^{4}V}\frac{\partial S}{\partial t}\left(  L_{3}S\right)
+\nonumber\\
&  \displaystyle+\frac{4}{r^{2}}\left(  \left(  L_{1}S\right)  ^{2}+\left(
L_{2}S\right)  ^{2}+\left(  L_{3}S\right)  ^{2}\right)  +\frac{4g^{2}a^{2}%
}{r^{6}V}\left(  L_{3}S\right)  ^{2}=-m^{2},
\end{align}
where we have defined $L_{i}=\xi_{i}^{R}$. The AdS signature is completely
encoded in the function $V$ and the last term of the left hand side,
proportional to $g^{2}$. This partial differential equation is completely
integrable, thanks to the symmetries of the metric (\ref{KSmetric}):
$\partial_{t}$, $\partial_{\alpha}$ and $\partial_{\gamma}$ are Killing
vectors of the spacetime. This suggests the ansatz
\begin{equation}
S=-Et+H(\alpha,\beta,\gamma)+W(r)\,,\qquad H(\alpha,\beta,\gamma)=j_{L}%
\alpha+j_{R}\gamma+\chi(\beta)\,. \label{HJansatz}%
\end{equation}
A further conserved quantity, $j^{2}$, arises from the St\"{a}ckel-Killing
tensor of the spacetime,
\begin{equation}
\left(  L_{1}H\right)  ^{2}+\left(  L_{2}H\right)  ^{2}+\left(  L_{3}H\right)
^{2}=j^{2}\,,
\end{equation}
which allows to determine $\chi$,
\begin{equation}
\frac{\partial\chi}{\partial\beta}=\pm\sqrt{j^{2}-\frac{1}{\sin^{2}\beta
}\left(  j_{R}^{2}+j_{L}^{2}-2j_{L}j_{R}\cos\beta\right)  }\,.
\end{equation}
Using the ansatz for the action, the Hamilton-Jacobi equation reduces to
\begin{equation}
W^{\prime}(r)^{2}=\frac{1}{V^{2}(r)}\left(  \Delta_{L}E^{2}-V\left(
m^{2}+\frac{4j^{2}}{r^{2}}\right)  -\frac{4\Delta}{r^{4}}aj_{R}E-\frac
{4g^{2}a^{2}}{r^{6}}j_{R}^{2}\right)  \,,
\end{equation}
which can be integrated.\newline The geodesic equations follow then from the
action,
\begin{equation}
\frac{dx^{\mu}}{d\lambda}=-g^{\mu\nu}\frac{\partial S}{\partial x^{\nu}}\,,
\end{equation}
where $\lambda$ denotes the geodesic parameter. For the radial motion, we
obtain
\begin{equation}
\left(  \frac{dr}{d\lambda}\right)  ^{2}=\Delta_{L}E^{2}-V\left(  m^{2}%
+\frac{4j^{2}}{r^{2}}\right)  -\frac{4\Delta}{r^{4}}aj_{R}E-\frac{4g^{2}a^{2}%
}{r^{6}}j_{R}^{2}\,. \label{geor}%
\end{equation}

One would like to know if geodesics can cross the velocity of light surface
and how far they can travel towards the horizon. Obviously, the motion is
allowed as long as the right hand side of equation (\ref{geor}) is positive.
Let us consider first geodesics with $j_{R}=0$. In this case we have
\begin{equation}
\dot{r}^{2}=\Delta_{L}E^{2}-V\left(  m^{2}+\frac{4j^{2}}{r^{2}}\right)
<\Delta_{L}E^{2},
\end{equation}
because $V>0$ for $r>r_{+}$. Moreover, $\Delta_{L}$ is positive outside the
velocity of light surface, but becomes negative for $r<r_{VLS}$. Hence
\emph{particles with $j_{R}=0$ cannot cross the VLS}. To enter the time
machine, we have to lower the potential barrier by putting some angular
momentum along the $\gamma$ direction. For $j_{R}\neq0$, we have two
additional terms in (\ref{geor}): the last term is always negative, but the
third term can become positive if $\Delta aj_{R}<0$. Hence, the geodesics can
enter the time machine only if they have spin \emph{opposite} to the dragging
effects of the spacetime.\newline Let us examine now if such a geodesic can
cross the event horizon. For $r=r_{+}$, the r.h.s.~of equation (\ref{geor})
becomes
\begin{equation}
\left.  \left(  \frac{dr}{d\lambda}\right)  ^{2}\right|  _{r_{+}}=\Delta
_{L}^{+}E^{2}-\frac{4\Delta_{+}}{r_{+}^{4}}aj_{R}E-\frac{4g^{2}a^{2}}%
{r_{+}^{6}}j_{R}^{2}\,, \label{geor+}%
\end{equation}
where $\Delta_{L}^{+}$ and $\Delta_{+}$ stand for the values on the horizon of
$\Delta_{L}$ and $\Delta$ respectively. The condition $V(r_{+})=0$ yields
\begin{equation}
\Delta_{L}^{+}=-\frac{\Delta_{+}^{2}}{g^{2}r_{+}^{2}}\,.
\end{equation}
Inserting this into (\ref{geor+}), we finally obtain
\begin{equation}
\left.  \left(  \frac{dr}{d\lambda}\right)  ^{2}\right|  _{r_{+}}=-\frac
{1}{g^{2}r_{+}^{2}}\left(  \Delta_{+}E+\frac{2g^{2}}{r_{+}^{2}}aj_{R}\right)
^{2}\leq0\,.
\end{equation}
If the previous quantity is strictly negative, it is obvious that the
geodesics cannot reach the horizon, but bounce back.
In the limiting case
\begin{equation}
E=-\frac{2g^{2}}{r_{+}^{2}\Delta_{+}}aj_{R}\,,
\end{equation}
the analysis needs more care. Writing the equation of motion (\ref{geor})
as $\dot r^2+U(r)=0$, we see that it describes the classical
one-dimensional motion of a particle in the potential $U(r)$. If
$U'(r_+) \neq 0$, then $r_+$ is a turning point, and the
geodesic bounces back without crossing the horizon.
In contrast, if $U'(r_+)$ vanishes, the particle needs an infinite time to
reach $r_+$, i. e. the geodesic approaches asymptotically the horizon for
$\lambda\rightarrow\infty$. But $\lambda$ is just the affine parameter, and
hence the geodesic ends there.
Note that this result still holds true for spacelike geodesics ($m^{2}<0$).

Hence, \emph{no geodesic can cross the horizon and penetrate the black hole
region in the spacetime} (\ref{KSmetric}). This means, moreover, that the
$r>r_{+}$ region of the manifold is \emph{geodesically complete}, and contains
no singularity\footnote{As a consequence, the solutions under consideration
are \emph{not} black holes. However, allowing for a slight abuse of language,
we shall continue to speak about ``black holes'' and ``horizons'' for
convenience.}.

\subsection{Charged geodesics}

We have seen that geodesics in our manifold cannot cross the horizon. We turn
now to the study of charged geodesics, and see if this property also holds for
charged particles. The Hamiltonian follows from the minimal coupling with the
$U(1)$ potential,
\begin{equation}
H=\frac{1}{2}g^{\mu\nu}\left(  p_{\mu}+QA_{\mu}\right)  \left(  p_{\nu
}+QA_{\nu}\right)  ,
\end{equation}
where $Q$ is the charge of the particle. The Hamilton-Jacobi equation can be
separated again by the ansatz (\ref{HJansatz}), and reads for charged
geodesics
\begin{align}
\left(  \frac{dr}{d\lambda}\right)^{2} =&\ \Delta_{L}E^{2}-V\left(  m^{2}+
\frac{4j^{2}}{r^{2}}\right)  -\frac{4\Delta}{r^{4}}aj_{R}E-\frac{4g^{2}a^{2}
}{r^{6}}j_{R}^{2}\nonumber\\
&  -6\Delta QE-\frac{12g^{2}a}{r^{2}}j_{R}Q-9\left(  \frac{g^{2}a^{2}}{r^{4}}
-\Delta^{2}\right)  Q^{2}\,. \label{georc}%
\end{align}
For $r=r_{+}$, the equation becomes
\begin{equation}
\left.  \left(  \frac{dr}{d\lambda}\right)  ^{2}\right|  _{r_{+}}=-\frac{1}{
g^{2}r_{+}^{2}}\left(  \Delta_{+}E+\frac{2g^{2}}{r_{+}^{2}} aj_{R}+3g^{2}%
r_{+}^{2}Q\right)  ^{2}<0\,.
\end{equation}
Being negative on the horizon, we can extend the previous conclusion to the
propagation of charged particles: \emph{charged geodesics cannot cross the
horizon of the black holes under consideration}.

\subsection{Scalar field propagation}

We consider now a neutral, minimally coupled scalar field propagating in the
background (\ref{KSmetric}). The Klein-Gordon equation
$\left(\nabla^{2}-m^{2}\right)\Phi=0$ reads
\begin{align}
-\frac{1}{r^{3}}\frac{\partial}{\partial r}\left(  r^{3}%
V\frac{\partial\Phi}{\partial r}\right) = &
-m^{2}\Phi-\frac{\Delta_{L}}{V}\frac{\partial^{2}\Phi}{\partial t^{2}}%
-\frac{4a\Delta}{r^{4}V}\frac{\partial}{\partial t}\left(  L_{3}\Phi\right)
\nonumber\\
&+\frac{4}{r^{2}}\left(  L_{1}^{2}+L_{2}^{2}+L_{3}^{2}\right)  \Phi
+\frac{4g^{2}a^{2}}{r^{6}V}L_{3}^{2}\Phi\,.
\end{align}
The variables can again be separated due to the symmetries of the problem. We
use the ansatz
\begin{equation}
\Phi=F(r)e^{-iEt}D_{j_{L},j_{R}}^{j}\,, \label{ansatz_wave}%
\end{equation}
with $D_{j_{L},j_{R}}^{j}$ the Wigner $D$-functions, which are simultaneous
eigenfunctions of $L_{3}$, $R_{3}$, $L^{2}$ and $R^{2}$:
\begin{align}
R_{3}D_{j_{L},j_{R}}^{j}&=-ij_{L}D_{j_{L},j_{R}}^{j},\quad
L_{3}D_{j_{L},j_{R}}^{j}=ij_{R}D_{j_{L},j_{R}}^{j},\\
L^{2}D_{j_{L},j_{R}}^{j}&=R^{2}D_{j_{L},j_{R}}^{j}%
=-j(j+1)D_{j_{L},j_{R}}^{j}.
\end{align}
We obtain then the radial wave equation,
\begin{equation}
-\frac{V}{r^{3}}\frac{d}{dr}\left(  r^{3}V\frac{dF}{dr}\right)  =\left[
\Delta_{L}E^{2}-V\left(  m^{2}+\frac{4j(j+1)}{r^{2}}\right)  -\frac{4a\Delta
}{r^{4}}Ej_{R}-\frac{4g^{2}}{r^{6}}a^{2}j_{R}^{2}\right]  F\,. \label{radeqn}%
\end{equation}
To eliminate the first order derivative from (\ref{radeqn}), we introduce a
tortoise-like coordinate
\begin{equation}
r_{\ast}(r)=\int\frac{dr}{r^{3}V(r)}\,, \label{tortoise}%
\end{equation}
which is a smooth strictly decreasing function of the radial coordinate $r$ in
the outer region; its range is $r_{\ast}\in]0,+\infty\lbrack$, where $r=r_{+}$
corresponds to $r_{\ast}\rightarrow+\infty$ and the spatial infinity
$r\rightarrow+\infty$ corresponds to $r_{\ast}=0$. In terms of this new
variable, the radial wave equation reads
\begin{equation}
\frac{d^{2}F}{dr_{\ast}^{2}}+U(r_{\ast})F=0\,, \label{radialeq}%
\end{equation}
where we have defined the potential function
\begin{equation}
U(r_{\ast})=r^{6}\left[  \Delta_{L}E^{2}-V\left(  m^{2}+\frac{4j(j+1)}{r^{2}%
}\right)  -\frac{4a\Delta}{r^{4}}Ej_{R}-\frac{4g^{2}}{r^{6}}a^{2}j_{R}%
^{2}\right]  \,.
\end{equation}
We are now interested in the behaviour of equation (\ref{radialeq}) near the
horizon (i.~e. for $r_{\ast}\rightarrow\infty$). To this end, we first observe
that the potential $U$ converges for $r\rightarrow\infty$; let us denote its
asymptotic value by $U_{0}$,
\begin{equation}
U_{0}\equiv\lim_{r_{\ast}\rightarrow\infty}U(r_{\ast})=-\frac{r_{+}^{4}}%
{g^{2}}\left(  \Delta_{+}E+\frac{2g^{2}aj_{R}}{r_{+}^{2}}\right)  ^{2}.
\label{U0}%
\end{equation}
Hence, performing a Taylor-Laurent expansion of $U(r_{\ast})$ near infinity,
the coefficients of the positive powers vanish, and we are left with
$U(r_{\ast})=U_{0}+\mathcal{O}\left(  r_{\ast}^{-1}\right)  $. The behaviour
near $r_{\ast}\rightarrow\infty$ of the solutions of the radial equation
(\ref{radialeq}) is determined by the sign of $U_{0}$: if it is positive, we
have oscillating solutions for $r\rightarrow r_{+} $, otherwise the
solutions are exponentially depressed or diverging. From equation
(\ref{U0}), we see that $U_{0}$ is always negative, and oscillating
solutions are therefore not possible. This means that the net flux of
particles through $r=r_{+}$ vanishes, and the total absorption cross
section of the horizon is zero:
\begin{equation}
\sigma_{abs}=0\,. \label{sigma_abs}%
\end{equation}
This result is exact and holds for any frequency. In fact it generalizes our
previous conclusions on geodesic motion, which is the high frequency, WKB
approximation. As a consequence of the non-existence of near-horizon
oscillating modes, there is no particle production and the Hawking temperature
vanishes, $T_{BH}=0$. Furthermore, as the absorption cross section
$\sigma_{abs}$ is a measure for the horizon area, (\ref{sigma_abs}) suggests
that one should assign zero entropy to the black holes (\ref{KSmetric}).

\section{General solution with vector supermultiplets}

\label{sect:vectmult}

The results obtained in the Einstein-Maxwell theory can be straightforwardly
generalized to $\mathcal{N}=2$, $D=5$ gauged supergravity coupled to $n$
vector supermultiplets. The bosonic part of the Lagrangian is given by
\begin{equation}
e^{-1}\mathcal{L}=\frac{1}{2}R+g^{2}\mathcal{W}-\frac{1}{4}G_{IJ}F_{\mu\nu
}^{I}F^{J\mu\nu}-\frac{1}{2}G_{IJ}\partial_{\mu}X^{I}\partial^{\mu}X^{J}%
+\frac{e^{-1}}{48}\epsilon^{\mu\nu\rho\sigma\lambda}C_{IJK}F_{\mu\nu}%
^{I}F_{\rho\sigma}^{J}A_{\lambda}^{K}, \label{genaction}%
\end{equation}
where $I=0,\ldots,n$. The scalar potential reads
\begin{equation}
\mathcal{W}(X)=\mathcal{W}_{I}\mathcal{W}_{J}\left(  6X^{I}X^{J}-\frac{9}%
{2}\mathcal{G}^{ij}\partial_{i}X^{I}\partial_{j}X^{J}\right)  . \label{pot}%
\end{equation}
Here $\mathcal{W}_{I}$ specify the appropriate linear combination of the
vectors that comprise the graviphoton of the theory, $\mathcal{A}_{\mu
}=\mathcal{W}_{I}A_{\mu}^{I}$. The $X^{I}$ are functions of the $n$ real
scalar fields, and obey the condition
\begin{equation}
\mathcal{V}=\frac{1}{6}C_{IJK}X^{I}X^{J}X^{K}=1.
\end{equation}
The gauge and the scalar couplings are determined in terms of the homogeneous
cubic polynomial $\mathcal{V}$ which defines a ``very special geometry''
\cite{antoine}. They are given by
\begin{align}
G_{IJ}  &  =-\frac{1}{2}\partial_{I}\partial_{J}\log\mathcal{V}\Big
|_{\mathcal{V}=1},\label{coupling}\\
\mathcal{G}_{ij}  &  =\partial_{i}X^{I}\partial_{j}X^{J}G_{IJ}\Big
|_{\mathcal{V}=1},\nonumber
\end{align}
where $\partial_{i}$ and $\partial_{I}$ refer, respectively, to partial
derivatives with respect to the scalar field $\phi^{i}$ and $X^{I}=X^{I}%
(\phi^{i})$.

For Calabi-Yau compactifications of M-theory, $\mathcal{V}$ denotes the
intersection form, and $X^{I}$ and $X_{I}\equiv\frac{1}{6}C_{IJK}X^{J}X^{K}$
correspond to the size of the two- and four-cycles of the Calabi-Yau threefold
respectively. Here $C_{IJK}$ are the intersection numbers of the threefold. In
the Calabi-Yau cases, $n$ is given by the Hodge number $h_{(1,1)}$.

The general charged rotating supersymmetric solution of the theory
(\ref{genaction}) reads \cite{ks2}
\begin{align}
ds^{2}&=-g^{2}r^{2}e^{2U}dt^{2}-e^{-4U}\left(  dt-\frac{\alpha}{r^{2}}
\sin^{2}\theta d\phi+\frac{\alpha}{r^{2}}\cos^{2}\theta d\psi\right)
^{2}+e^{2U}\left(\frac{dr^{2}}{V(r)}+r^{2}d\Omega_{3}^{2}\right),
\nonumber\\
A_{\phi}^{I}  &  =-e^{-2U}X^{I}\frac{\alpha}{r^{2}}\sin^{2}\theta,\quad
A_{\psi}^{I}=e^{-2U}X^{I}\frac{\alpha}{r^{2}}\cos^{2}\theta,\quad A_{t}%
^{I}=e^{-2U}X^{I}, \label{gensol}%
\end{align}
where
\begin{equation}
V(r)=1+g^{2}r^{2}e^{6U}-\frac{g^{2}\alpha^{2}}{r^{4}}.
\end{equation}
As a particular case, one obtains for the $STU=1$ model\footnote{We apologize
for using the same symbol for one of the moduli and the function appearing in
the metric, but the meaning should be clear from the context.} ($X^{0}=S$,
$X^{1}=T$, $X^{2}=U$)
\begin{equation}
e^{6U}=H_{1}H_{2}H_{3}\,,\qquad H_{I}=h_{I}+\frac{q_{I}}{r^{2}}\,,\quad
I=0,1,2, \label{UH}%
\end{equation}
and
\begin{equation}
X^{I}=e^{2U}H_{I}^{-1}.
\end{equation}
Taking all charges to be equal, $q_{I}=q$, and $h_{I}=1$ for $I=0,1,2$, the
solution of the $STU=1$ model reduces to the Einstein-Maxwell solution
(\ref{KSmetric}) considered in the previous sections. Using Euler coordinates
on the three-sphere and the vielbeins (\ref{sigma}), the general metric
(\ref{gensol}) can be cast into the form
\begin{equation}
ds^{2}=-g^{2}r^{2}e^{2U}dt^{2}-e^{-4U}\left(  dt+\frac{\alpha}{2r^{2}}%
\ \sigma^{3}\right)  ^{2}+e^{2U}\left(  \frac{dr^{2}}{V(r)}+r^{2}d\Omega
_{3}^{2}\right)  . \label{generalmetric}%
\end{equation}
``Horizons'' occur for $r=r_{+}$ with $V(r_{+})=0$, while the VLS is located
at the zero $r_{VLS}$ of the function $\Delta_{L}(r)$, defined by
\begin{equation}
\Delta_{L}(r)=1-\frac{\alpha^{2}}{r^{6}e^{6U}}\,. \label{DeltaL}%
\end{equation}
It is straightforward to show that $V(r)>g^{2}r^{2}e^{6U}\Delta_{L}(r)$, from
which it follows that the VLS is always external to the horizon,
$r_{+}<r_{VLS}$, as in the Einstein-Maxwell case: there are always naked
closed timelike curves.

\subsection{Geodesic motion and Scalar field propagation}

To see if the general solution represents a black hole, we have to study its
maximal extension. The Hamilton-Jacobi equation for the geodesic motion is
\begin{align}
&  \displaystyle-\frac{\Delta_{L}}{V}e^{4U}\left(  \frac{\partial S}{\partial
t}\right)  ^{2} +Ve^{-2U}\left(  \frac{\partial S}{\partial r}\right)  ^{2}
-\frac{4\alpha}{r^{4}V}e^{-2U}\frac{\partial S}{\partial t} \left(
L_{3}S\right)  +\nonumber\\
&  \displaystyle+\frac{4}{r^{2}}e^{-2U}\left(  \left(  L_{1}S\right)
^{2}+\left(  L_{2}S\right)  ^{2}+\left(  L_{3}S\right)  ^{2}\right)
+\frac{4g^{2}\alpha^{2}}{r^{6}V}e^{-2U}\left(  L_{3}S\right)  ^{2}=-m^{2}.
\end{align}
Again, the Hamilton-Jacobi equation can be completely separated using the
ansatz (\ref{HJansatz}). The orbital motion is the same as for the
Einstein-Maxwell case, while the radial equation of motion becomes
\begin{equation}
\left(  \frac{dr}{d\lambda}\right)  ^{2}=e^{-4U}\left[  e^{6U}\Delta_{L}%
E^{2}-V\left(  m^{2}e^{2U}+\frac{4j^{2}}{r^{2}}\right)  -\frac{4\alpha}{r^{4}%
}j_{R}E-\frac{4g^{2}\alpha^{2}}{r^{6}}j_{R}^{2}\right]  \,. \label{geor_gen}%
\end{equation}
To see if the geodesics can cross the horizon, we compute the {r.h.s.}~of
Eq.~(\ref{geor_gen}) for $r=r_{+}$, with $r_{+}$ a zero of the function
$V(r)$. This yields
\begin{equation}
\left.  \left(  \frac{dr}{d\lambda}\right)  ^{2}\right|  _{r_{+}}
=-\frac{e^{-4U(r_{+})}}{g^{2}r_{+}^{2}}\left(  E+\frac{2g^{2}\alpha}{r_{+}%
^{2}}j_{R}\right)  ^{2} \leq0\,.
\end{equation}
This is negative, and applying the argument of section \ref{sect:geodesics} we
can conclude that no geodesic can cross the $r=r_{+}$ hypersurface. Hence, the
region $r>r_{+}$ is geodesically complete and non-singular.

The Klein-Gordon equation for a neutral, minimally coupled scalar field of
mass $m$ propagating in the background of the general solution
(\ref{generalmetric}) reads
\begin{align}
-\frac{e^{-2U}}{r^{3}}\frac\partial{\partial r} \left(  r^{3}V\frac
{\partial\Phi}{\partial r}\right) = &
-m^{2}\Phi
-e^{4U}\frac{\Delta_{L}}V\frac{\partial^{2}\Phi}{\partial t^{2}}
-e^{-2U}\frac{4\alpha}{r^{4}V}\frac\partial{\partial t}
\left(  L_{3}\Phi\right) \nonumber\\
&  +\frac{4e^{-2U}}{r^{2}}\left(  L_{1}^{2}+L_{2}^{2}+L_{3}^{2}\right)\Phi
+\frac{4g^{2}\alpha^{2}}{r^{6}V}e^{-2U}L_{3}^{2}\Phi\,.
\end{align}
Making the ansatz (\ref{ansatz_wave}), the variables separate, leaving the
radial wave equation
\begin{eqnarray}
\lefteqn{-\frac1{r^{3}}\frac{d}{dr}\left(  r^{3}V\frac{dF}{dr}\right)  =}
\nonumber \\
&&\left[
e^{6U}\frac{\Delta_{L}}VE^{2}-m^{2}e^{2U}-\frac{4j(j+1)}{r^{2}} -\frac
{4\alpha}{r^{4}V}Ej_{R}-\frac{4g^{2}\alpha^{2}}{r^{6}V}j_{R}^{2}\right]  F\,.
\end{eqnarray}
Using the Regge-Wheeler coordinate $r_{\ast}$ defined in (\ref{tortoise}), we
obtain the differential equation
\begin{equation}
\frac{d^{2}F}{dr_{\ast}^{2}}+P(r_{\ast})F=0\,, \label{radialeq2}%
\end{equation}
describing a classical particle moving in the potential
\begin{equation}
P(r_{\ast})=r^{6}\left[  e^{6U}\Delta_{L}E^{2}-V\left(  m^{2}e^{2U}%
+\frac{4j(j+1)}{r^{2}}\right)  -\frac{4\alpha}{r^{4}}Ej_{R}-\frac{4g^{2}%
\alpha^{2}}{r^{6}}j_{R}^{2}\right]  \,.
\end{equation}
For $r=r_{+}$ ($r_{\ast}\rightarrow\infty$), the potential converges to the
finite value $P_{0}$,
\begin{equation}
P_{0}\equiv\lim_{r_{\ast}\rightarrow\infty}P(r_{\ast})= -\frac{r_{+}^{4}%
}{g^{2}}\left(  E+\frac{2g^{2}\alpha}{r_{+}^{2}}j_{R}\right)  ^{2}\,,
\label{P0}%
\end{equation}
which is always negative. Hence, as shown previously, the total flux across
the $r=r_{+}$ hypersurfaces is zero, and the total absorption cross section
vanishes, $\sigma_{abs}=0$. These results suggest to assign zero temperature
and entropy also to the general solution (\ref{generalmetric}).

\section{Lifting to type IIB supergravity}

\label{sect:lift}

Let us now focus on the $STU=1$ model, with scalar potential (\ref{pot}) and
gauge couplings $G_{IJ}$ (\ref{coupling}) given by \cite{bcs}\footnote{We
assumed $h_{I}=1$, yielding $\mathcal{W}_{I}=\frac{1}{3}$ \cite{ks2}.}
\begin{align}
\mathcal{W}  &  =2\left(  \frac{1}{U}+\frac{1}{T}+TU\right)  ,\\
G_{IJ}  &  =\frac{1}{2}\left(
\begin{array}
[c]{ccc}%
T^{2}U^{2} & 0 & 0\\
0 & \frac{1}{T^{2}} & 0\\
0 & 0 & \frac{1}{U^{2}}%
\end{array}
\right)  .
\end{align}
In \cite{tenauthors} it was shown that this model can be obtained from type
IIB supergravity in ten dimensions by the Kaluza-Klein reduction ansatz
\begin{equation}
ds_{10}^{2}=\sqrt{\tilde{\Delta}}ds_{5}^{2}+\frac{1}{g^{2}\sqrt{\tilde{\Delta
}}}\sum_{I=0}^{2}(X^{I})^{-1}\left(  d\mu_{I}^{2}+\mu_{I}^{2}(d\phi_{I}%
+gA^{I})^{2}\right)  , \label{redansatz}%
\end{equation}
where the three quantities $\mu_{I}$ are subject to the constraint $\sum
_{I}\mu_{I}^{2}=1$. The standard metric on the unit five-sphere can be written
in terms of these as
\begin{equation}
d\Omega_{5}^{2}=\sum_{I}(d\mu_{I}^{2}+\mu_{I}^{2}d\phi_{I}^{2}).
\end{equation}
The $\mu_{I}$ can be parametrized in terms of angles on a two-sphere,
e.~g.~as
\begin{equation}
\mu_{0}=\sin\Theta,\qquad\mu_{1}=\cos\Theta\sin\Psi,\qquad\mu_{2}=\cos
\Theta\cos\Psi.
\end{equation}
$\tilde{\Delta}$ is given by
\begin{equation}
\tilde{\Delta}=\sum_{I=0}^{2}X^{I}\mu_{I}^{2}.
\end{equation}
The ansatz for the reduction of the 5-form field strength is $F_{(5)}%
=G_{(5)}+\ast G_{(5)}$, where \cite{tenauthors}
\begin{align}
G_{(5)} = &\  2g\sum_{I}\left(  (X^{I})^{2}\mu_{I}^{2}-\tilde{\Delta}%
X^{I}\right)  \epsilon_{(5)}-\frac{1}{2g}\sum_{I}(X^{I})^{-1}\bar{\ast}%
dX^{I}\wedge d(\mu_{I}^{2})\nonumber\\
&  +\frac{1}{2g^{2}}\sum_{I}(X^{I})^{-2}d(\mu_{I}^{2})\wedge(d\phi_{I}%
+gA^{I})\wedge\bar{\ast}dA^{I}\,,
\end{align}
here $\epsilon_{(5)}$ is the volume form of the five-dimensional metric
$ds_{5}^{2}$, and $\bar{\ast}$ denotes the Hodge dual with respect to the
five-dimensional metric $ds_{5}^{2}$.

The other bosonic fields of the type IIB theory are zero in this $U(1)^{3}$
truncated reduction.

Using (\ref{redansatz}), we can lift our charged rotating supersymmetric
solutions (\ref{generalmetric}), with $U$ given by (\ref{UH}), to ten
dimensions, obtaining
\begin{align}
ds_{10}^{2} = & \sqrt{\tilde{\Delta}}\left[  -g^{2}r^{2}e^{2U}dt^{2} -
e^{-4U} \left(  dt + \frac{\alpha}{2r^{2}}\sigma^{3}\right)  ^{2} +
e^{2U}\left(  \frac{dr^{2}}{V(r)} + r^{2}d\Omega_{3}^{2}\right)  \right]
\nonumber\\
&  + \frac{1}{g^{2}\sqrt{\tilde{\Delta}}}\sum_{I=0}^{2} (X^{I})^{-1}\left(
d\mu_{I}^{2} + \mu_{I}^{2}(d\phi_{I} + gH_{I}^{-1}(dt + \frac{\alpha}{2r^{2}}
\sigma^{3}))^{2}\right)  . \label{10dmetric}%
\end{align}
(\ref{10dmetric}) represents a D3-brane rotating both in directions transverse
and longitudinal to the world volume.

Note that in five dimensions, the norm squared of the Killing vector
$\partial_{\gamma}$ is
\begin{equation}
(\partial_{\gamma},\partial_{\gamma})_{(5)}=e^{2U}\frac{r^{2}}{4}\Delta
_{L}(r), \label{5dnorm}%
\end{equation}
where $\Delta_{L}(r)$ is given by Eq.~(\ref{DeltaL}). As we said, for
$r<r_{VLS}$ (\ref{5dnorm}) becomes negative, so we have closed timelike
curves. We would like to see whether these CTCs still occur in ten dimensions.
A straightforward calculation shows that the norm squared of $\partial
_{\gamma}$ computed with the ten-dimensional metric (\ref{10dmetric}) reads
\begin{equation}
(\partial_{\gamma},\partial_{\gamma})_{(10)}=\sqrt{\tilde{\Delta}}e^{2U}%
\frac{r^{2}}{4},
\end{equation}
which is always positive. Thus in the ten-dimensional metric $\partial
_{\gamma}$ is always spacelike, and the CTCs present in five dimensions
disappear upon lifting. However, new CTCs show up. Consider e.~g.~the vector
\begin{equation}
v:=-\frac{e^{-2U}\alpha g}{2r^{2}}\sum_{I=0}^{2}\partial_{\phi_{I}}%
+\partial_{\gamma},
\end{equation}
which has closed orbits, and norm
\begin{equation}
v^{2}=e^{2U}\frac{r^{2}}{4}\Delta_{L}(r),
\end{equation}
that becomes negative for $r<r_{VLS}$. Thus in our case we have no resolution
of causal anomalies in higher dimensions, in contrast to the BMPV black hole
\cite{herd}.\newline The naively computed Bekenstein-Hawking entropy of the
rotating D3 brane (\ref{10dmetric}) reads
\begin{equation}
S=\frac{V_{3}V_{5}}{4g^{5}G_{10}}\sqrt{e^{6U(r_{+})}r_{+}^{6}-\alpha^{2}},
\label{Snaive}%
\end{equation}
where $V_{3}$ and $V_{5}$ denote the volume of the unit $\mathcal{S}^{3}$ and
$\mathcal{S}^{5}$ respectively. (\ref{Snaive}) is clearly imaginary and thus
makes no sense.

\section{Final remarks}

\label{sect:finrem}

The solutions (\ref{KSmetric}) and (\ref{gensol}) that we studied in this
paper are BPS states preserving half of the supersymmetry of $\mathcal{N}=2$,
$D=5$ gauged supergravity. The fact that we obtained a vanishing Hawking
temperature $T_{H}$ due to the non-existence of near-horizon oscillating modes
is in agreement with the supersymmetry considerations. Note that we obtained
$T_{H}=0$ also for solutions with a simple root $r_{+}$ of the function
$V(r)$. One would expect for such solutions a bifurcated horizon, and a finite
Hawking temperature. The resolution of this puzzle is related to the bad
causal behaviour of the spacetimes under consideration.

Another question to settle, is whether it is possible to construct and use
such time machines. For example, one could take (for simplicity) the solution
with $q=0$ and $a=0$, which is AdS$_{5}$, and try to add angular momentum by
turning on $a$. This leads to CTCs outside the horizon. But this process,
which is similar to accelerating a particle beyond the speed of light
\cite{gibbherd}, should be impossible due to Hawking's chronology protection
conjecture \cite{cronprot}.
We expect furthermore, in analogy with the ungauged case \cite{herd}, that
these causality-violating solutions should be forbidden, because they would
correspond to states which violate the unitarity bound of the dual
$\mathcal{N}=4$, $D=4$ SYM theory.
In such a way, the AdS/CFT correspondence would provide us with a nice
implementation of the chronology protection conjecture in AdS spacetimes.

This effect already shows up for the uncharged nonextremal
Kerr-AdS$_5$ black holes found in \cite{Hawking:1999kw}, where CTCs
appear outside
the horizon if the two rotation parameters are equal, and if the mass
parameter $M$ is sufficiently negative. From the dual CFT point of view,
it is clear that these CTCs are related to loss of unitarity.
The reason is that the classification of unitary representations of
superconformal algebras typically implies inequalities on the conformal
weights and R-charges. These inequalities thus yield lower bounds on the
black hole mass $M$, and therefore the unitarity bound of the
superconformal algebra is violated if $M$ becomes too negative.
It remains to be shown that this unitarity bound exactly coincides with
the point in parameter space where CTCs develop.

\acknowledgments
D.~K.~was partially supported by MURST and by the European Commission RTN
program HPRN-CT-2000-00131, in which he is associated to the University of
Torino. The authors would like to thank Klaus Behrndt for useful discussions.

\end{document}